# Unexpected Suppression of Leidenfrost Phenomenon on Superhydrophobic Surfaces

Meng Shi, Ratul Das, Sankara Arunachalam, Himanshu Mishra*

Interfacial Lab (iLab), Water Desalination and Reuse Center (WDRC),
Biological and Environmental Science and Engineering (BESE) Division,
King Abdullah University of Science and Technology (KAUST),
Thuwal, 23955-6900, Saudi Arabia

*Corresponding author: himanshu.mishra@kaust.edu.sa

**Abstract**

The Leidenfrost phenomenon entails the levitation of a liquid droplet over a superheated surface, cushioned by its vapor layer. For water, superhydrophobic surfaces are believed to suppress the Leidenfrost point ($T_L$)—the temperature at which this phenomenon occurs. The vapor film obstructs boiling heat transfer in heat exchangers, thereby compromising energy efficiency and safety. Thus, it is desirable to realize superhydrophobicity without suppressing $T_L$. Here we demonstrate that the $T_L$ of water on microtextured superhydrophobic surfaces comprising doubly reentrant pillars (DRPs) can exceed those on hydrophilic and even superhydrophilic surfaces. We disentangle the contributions of microtexture, heat transfer, and surface chemistry on $T_L$ and reveal how superhydrophobicity can be realized without suppressing $T_L$. For instance, silica surfaces with DRPs facilitate ~300% greater heat transfer to water droplets at 200°C in comparison with silica surfaces coated with perfluorinated-nanoparticles. Thus, superhydrophobic surfaces could be harnessed for energy efficient thermal machinery.

**Introduction**

Known for over two centuries, the Leidenfrost phenomenon entails the levitation of an evaporating liquid droplet on a hot surface such that the droplet's weight is counterbalanced by the pressure of the vapor film formed underneath[1]. This phenomenon has attracted interest across many disciplines[2, 3, 4, 5, 6, 7, 8, 9, 10, 11, 12], and it has been applied to engineering applications, such as directional droplet propulsion[7, 9], electricity generation[13], and green chemistry[14], among others[3]. For liquids with strong intermolecular forces[15] and a high boiling point such as water, the Leidenfrost phenomenon occurs—the Leidenfrost point ($T_L$)—at temperatures significantly higher than at normal temperature and pressure (NTP: 293 K and 1 atm). For example, for a smooth and flat silica surface, the $T_L$ of water is in the range 270°C–300°C[16, 17]. As a corollary, the Leidenfrost phenomenon can be realized at NTP with liquids that have ultralow boiling points, such as liquid nitrogen and 1,1,1,2-tetrafluoroethane with boiling points of −196°C[18] and −26.1°C[9], respectively. In either scenario, a vapor film with a thickness of ~100 μm levitates the droplet above the surface, after which the droplet slowly evaporates, akin to film boiling[3, 9, 18]. Roughening of surfaces can have profound consequences on $T_L$: for example, hierarchical micro/nano-texturing furthers the wettability of a silica surface with water, increasing the temperature needed for the water droplet to achieve the Leidenfrost state (e.g., $T_L$ > 400°C)[16, 17, 19, 20]. Conversely, when hydrophobic coatings are applied to micro/nano-textured surfaces, they can give rise to superhydrophobicity, characterized by water contact angles, $\theta_r \geq 150°$, and an empirical constraint on contact angle hysteresis, e.g., $\Delta\theta \leq 20°$[21, 22, 23]. Superhydrophobicity lowers the $T_L$ of water by facilitating (i) low adhesion energy per unit area at the solid–liquid–vapor (SLV) interface given by the Young-Dupré equation, $W_{ad} = \gamma_{LV} \times (1 + \cos\theta_o)$, where $\gamma_{LV}$ is the surface tension, and $\theta_o > 90°$ is the actual (or intrinsic) contact angle of the liquid (water)[24], and (ii) reduced liquid–solid contact area (hence adhesion) due to the interfacial entrapment of air[4, 25, 26, 27, 28]. For instance, after a flat metallic surface was rendered superhydrophobic by coating it with the Glaco Mirror Coat Zero™ coating (hereafter referred to as Glaco), the $T_L$ of water reduced from 210°C to 130°C; similarly, Glaco-coated superhydrophobic steel balls[4] and rotors[26] exhibited $T_L$ approaching 100°C and also afforded significant reduction in frictional drag[4]. Taken together, these findings give credence to conventional wisdom that superhydrophobicity lowers the $T_L$ of water [25, 27, 28, 29].

Superhydrophobic surfaces have been exploited to enhance pool boiling[30]; however, their applications in thermal machinery operating at high temperatures remain limited as superhydrophobicity generally lowers the $T_L$ of water. Specifically, the vapor film prevents the interfacial heat transfer because of its lower thermal conductivity than that of the liquid[31], for example, in boiler pipes, heat exchangers, and turbines. This situation, known as boiling crisis or burnout[32], can severely impact energy efficiency, damage machinery, and cause accidents[33]. Hence, it is of practical interest to investigate whether the lowering of the $T_L$ of water is a universal attribute of superhydrophobicity, similarly to the water contact angle, $\theta_r > 150°$, the contact angle hysteresis, $\Delta\theta < 20°$, and the robust entrapment of air under water on superhydrophobic surfaces, because superhydrophobicity without suppressing the $T_L$ of water could reduce frictional drag[34] while facilitating interfacial heat transfer.

Here, contrary to conventional wisdom, we demonstrate that superhydrophobic surfaces do not always lower the $T_L$ of water. We utilize silica surfaces adorned with a specialized biomimetic microtexture comprising doubly reentrant pillars (DRPs) that exhibit superhydrophobicity despite their hydrophilic surface chemistry.[22, 23, 35, 36] Counterintuitively, water's $T_L$ on these superhydrophobic silica surfaces is significantly higher than that on flat hydrophilic silica surfaces and even higher than that on superhydrophilic microtextured silica surfaces (see Table 1 for the apparent contact angles of water droplets). We combine experiment and theory to pinpoint the contributions of microtexture, interfacial heat transfer, and surface chemistry to the onset of the Leidenfrost phenomenon on these surfaces.

**Results**

We utilized $SiO_2$/Si wafers and Glaco-coated $SiO_2$/Si wafers as hydrophilic and superhydrophobic surfaces (hereafter referred to as flat silica and Glaco-coated silica, respectively) **(Figs. 1a and 1b)**. Additionally, we microfabricated arrays of cylindrical pillars **(Fig. 1c)** and DRPs with diameter, $D$ = 10 μm, height, $H$ = 50 μm, and pitch, $L$ = 50 μm **(Fig. 1d-f)**. The details of the microfabrication processes have been reported recently; briefly, silicon wafers with a 2.4-μm-thick silica layer were prepared by photolithography, dry etching, and thermal oxide growth[22, 23, 37, 38] **(Methods)**.

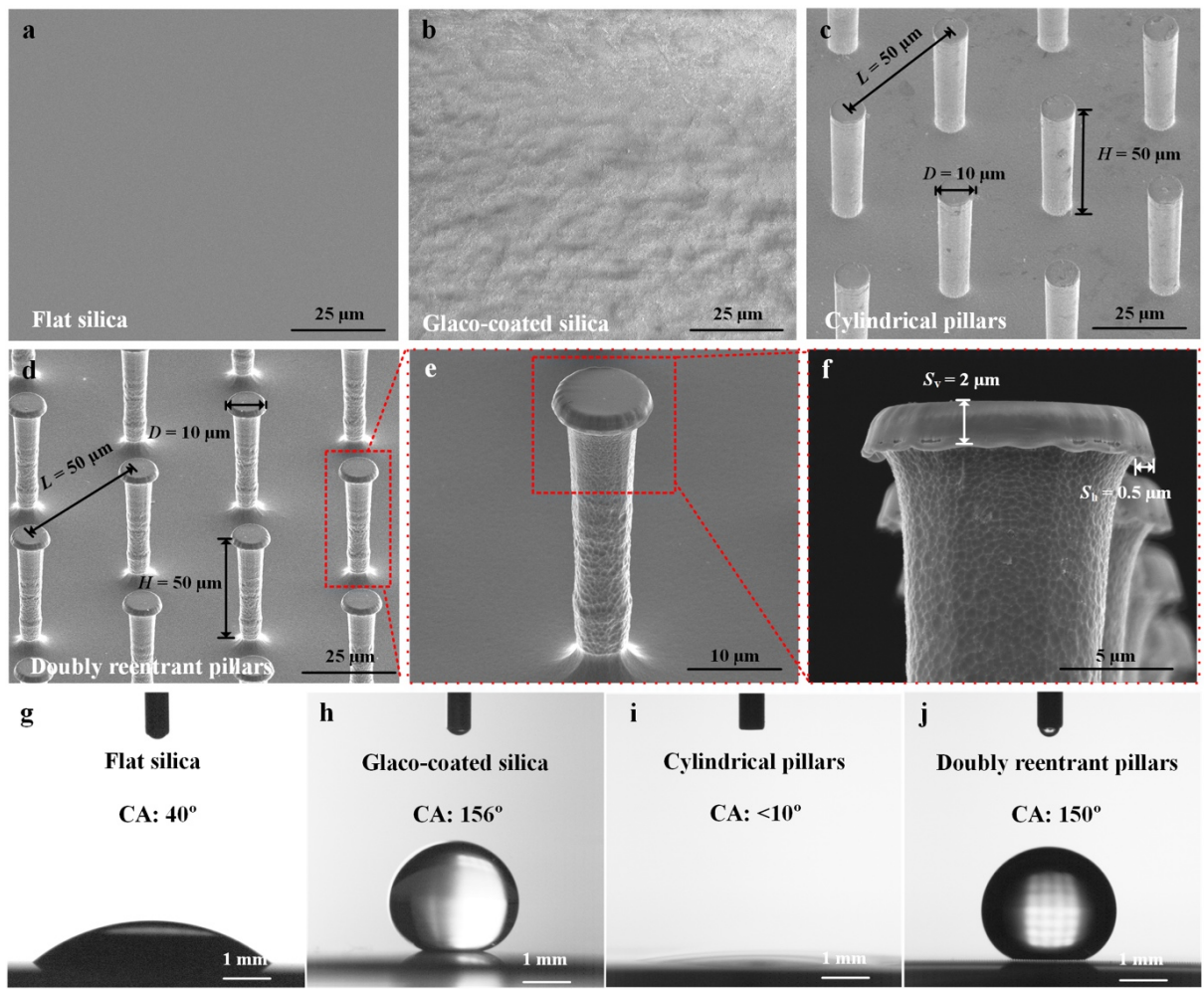

**Fig. 1| Representative scanning electron micrographs of surfaces investigated in this study and their wettability with water at NTP.** a) Flat silica; b) Glaco-coated silica that exhibits superhydrophobicity; c) tilted view (45°) of cylindrical pillars with a diameter (*D*) of 10 μm, height (*H*) of 50 μm, and pitch (*L*) of 50 μm; d) tilted view (45°) of doubly reentrant pillars with *D* = 10 μm, *H* = 50 μm, and *L* = 50 μm; e, f) magnified views of DRPs with the vertical length and width of the double reentrant edge, respectively, $S_v$ = 2 μm and $S_h$ = 0.5 μm. g–j) Representative images of water droplets (15 μL) on the surfaces outlined in panels (a–d) and their apparent contact angles (CAs).

First, we characterized the wettability of these surfaces with water by measuring the advancing ($\theta_A$) and receding ($\theta_R$) contact angles (CAs) of droplets (6 μL) advanced/retracted at 0.2 μL/s **(Methods)**. On flat silica, $\theta_A$ = 45° ± 2°, $\theta_R$ = 0°, and $\theta_r$ = 40° ± 2° (**Fig. 1g**); therefore we considered 40° to be the actual (or intrinsic) contact angle ($\theta_o$) for silica material in our theoretical analysis[21]. Glaco-coated silica exhibited superhydrophobicity with $\theta_r$ = 156° ± 1° and contact angle hysteresis $\Delta\theta = \theta_A - \theta_R$ =19° ± 2° (**Fig. 1h**). Conversely, silica surfaces with cylindrical pillars were superhydrophilic—they imbibed water, thus reaching the fully filled (or Wenzel) state[39] (**Fig. 1i**); apparent CAs of water droplets were ultralow, $\theta_r$ < 10°. Silica surfaces with DRPs exhibited superhydrophobicity with $\theta_r$ = 150° ± 1° and $\Delta\theta$ =19° ± 2° (**Fig. 1j**), similarly to Glaco-coated silica. The sliding angles of water droplets (10 μL) on Glaco-coated silica and silica with DRPs were 7° ± 3° and 13° ± 2°, respectively (**Table 1 and Methods**).

**Table 1. Characterization of various surfaces investigated in this work**

| Substrates | Apparent contact angles | | | Contact angle hysteresis | Nature of the substrate | Sliding angle | Measured Leidenfrost point, $T_L$ (°C) |
|---|---|---|---|---|---|---|---|
| | $\theta_R$ | $\theta_r$ | $\theta_A$ | $\Delta\theta$ | | | $T_L$ |
| Flat silica | 0° | 40° ± 2° | 45° ± 2° | 45° ± 2° | Hydrophilic | / | ~286 |
| Glaco-coated silica | 151° ± 3° | 156° ± 1° | 170° ± 1° | 19° ± 2° | Superhydrophobic | 7° ± 3° | ~130 |
| Cylindrical pillars | 0° | <10° | <10° | <10° | Superhydrophilic | / | ~314 |
| DRPs | 146° ± 2° | 150° ± 1° | 165° ± 2° | 19° ± 2° | Superhydrophobic | 13° ± 2° | ~363 |

Next, we investigated the response of water droplets on the following substrates placed on a leveled hot plate to control the surface temperature precisely (±1°C): (i) flat silica, (ii) Glaco-coated silica, and silica surfaces with arrays of (iii) cylindrical pillars and (iv) DRPs. We arbitrarily chose an initial temperature of 347°C, which is beyond the $T_L$ of water on common surfaces[40]. Water droplets (15 μL) were gently placed onto the surfaces, and the subsequent behaviors were recorded via high-speed imaging. The placement of droplets ensured that the Weber numbers, defined as the ratio of the droplet's inertia to capillarity, $We = \rho V^2 R_d / \gamma_{LV}$, where $\rho$, $R_d$, $\gamma_{lv}$, and $V$ is the density, radius, water surface tension, and impact velocity, respectively, were as small as possible (~1.7).[3] Contrary to our expectation, all samples except superhydrophobic DRPs exhibited the Leidenfrost phenomenon at 347°C (**Supplementary Materials, Movies S1–S4, Figs. 2a–d**). Specifically, after the initial impact and followed by 3-6 bounces due to inertia, water droplets re-established contact with the DRPs with slight vibrations (**Supplementary Materials, Movie S4**).

To confirm this result, we ascertained $T_L$ on these surfaces by tracking the lifetime of sessile water droplets on them as a function of surface temperature to determine how long it took for the water droplets to evaporate on these surfaces. As the surface temperature ($T_s$ (°C)) increases, the droplet lifetime decreases due to faster evaporation. In the vicinity of $T_L$, the droplet lifetime starts increasing monotonically due to the formation of an insulating vapor layer underneath the droplet that diminishes the interfacial heat transfer. By gradually increasing the temperature in the range 70°C–440°C, the $T_L$ values for hydrophilic silica

surfaces—flat and those adorned with cylindrical pillars—were pinpointed to be ~286°C and ~314°C, respectively, (**Fig. 2e**).

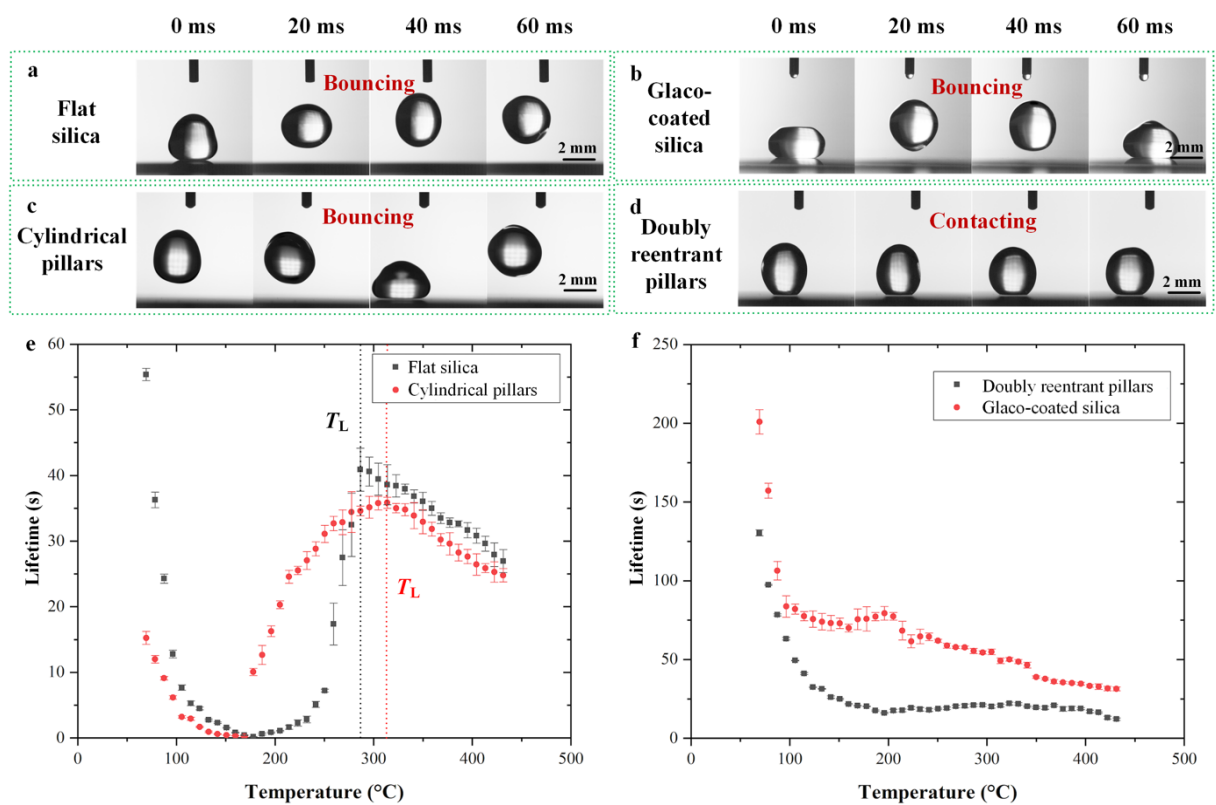

**Fig. 2| High-speed micrographs of water droplets exhibiting Leidenfrost phenomena after gentle placement on the following surfaces maintained at 347°C**: a) flat silica (hydrophilic); b) Glaco-coated silica (superhydrophobic); c) silica surface with cylindrical pillars with $D = 10$ μm, $H = 50$ μm, and $L = 50$ μm (superhydrophilic); and d) silica surfaces with doubly reentrant pillars (DRPs) with $D = 10$ μm, $H = 50$ μm, and $L = 50$ μm (superhydrophobic). These snapshots are derived from **Supplementary Movies S1–S4**. Lifetimes of water droplets placed on the surfaces described in (a–d) as a function of surface temperature: e) hydrophilic flat silica surface and superhydrophilic silica surfaces with cylindrical pillars and f) superhydrophobic surfaces (Glaco-coated silica and silica with DRPs).

For superhydrophobic surfaces—DRPs and Glaco-coated silica—there were no distinct peaks in the droplet lifetime curves **(Fig. 2f)**. Therefore, we utilized high-speed imaging to monitor vapor film formation at the droplet-solid interface to pinpoint the onset of Leidenfrost phenomena (**Supplementary Materials, Figs. S3** and **S4**). For water droplets, superhydrophobic Glaco-coated silica exhibited $T_L$ ~130°C. In stark contrast, however, superhydrophobic silica surfaces with DRPs did not exhibit the Leidenfrost phenomenon until the surface temperature was raised to ~363°C. This dramatic variation in the $T_L$ values of superhydrophobic surfaces challenges our current understanding of the factors and mechanisms underlying the Leidenfrost phenomenon.

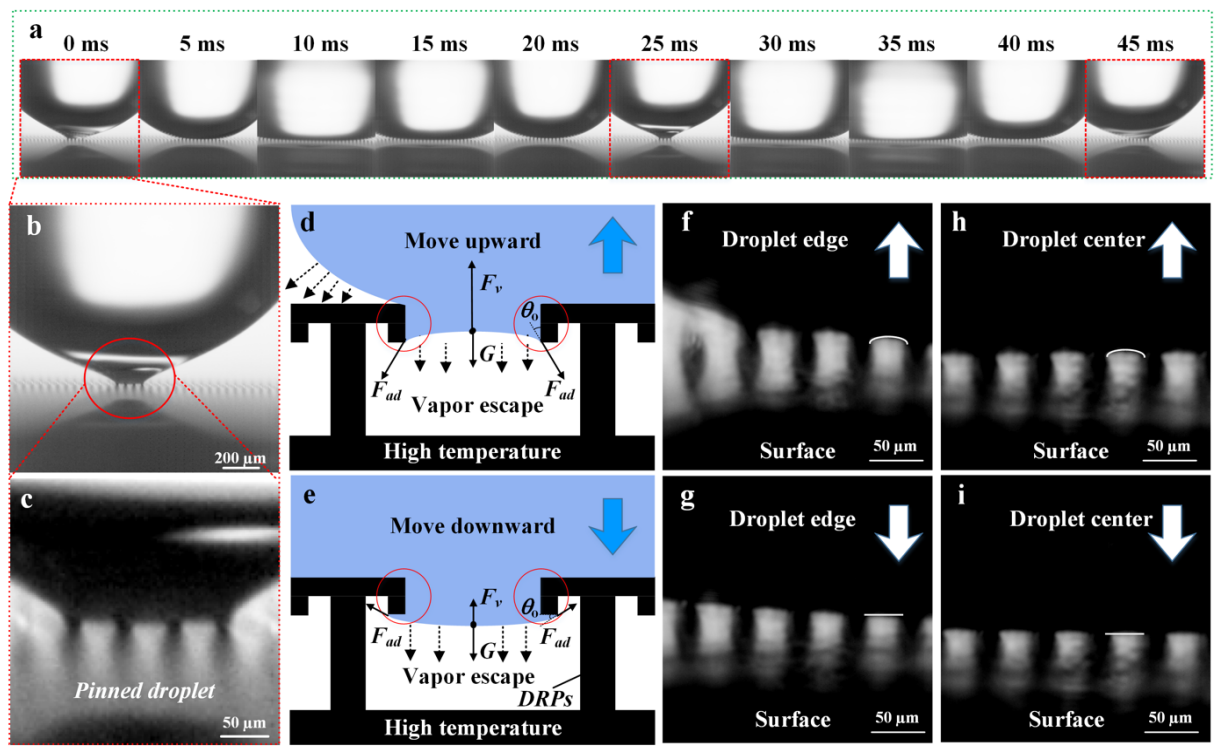

**Fig. 3| Pinning of water droplets on silica surfaces adorned with DRPs maintained at 347°C**. a) The water droplet is unable to detach completely from the DRPs, and it exhibits an oscillatory motion during which it contacts a varying number of pillars; b, c) droplet at its furthest point from the surface, yet pinned on a few DRPs. Schematics of the forces at the liquid–solid–vapor interface due to adhesion ($F_{Ad}$), weight ($G$), and vapor force ($F_v$) as the drop moves d) upward and e) downward. f–i) High-speed snapshots of the curvature of the droplet–DRPs interface moving f–h) upward and g–i) downward.

Next, we observed the behavior of water droplets (15 μL) on silica with DRPs at $T = 347°C$ via high-speed imaging complemented with high-magnification optics (**Fig. 3**). We found that a typical droplet exhibited an up-and-down vibratory motion under the influence of its vapor force built underneath, weight, and adhesion at the SLV interface (**Figs. 3a–3c**, and **Supplementary Materials, Video S5**). As the droplet moved upward, the number of DRPs in contact with the droplet decreased and vice versa, and the air–water interface assumed concave and convex curvatures, respectively (**Figs. 3d and 3e**). After the initial contact, when the center of mass of the droplet was at the highest point, it was still pinned to ~3–10 DRPs, which preempted the onset of the Leidenfrost phenomenon (**Figs. 3b and 3c**). We considered that as a liquid droplet is placed on a hot substrate, its vapor flow is obstructed by the surface microtexture, which builds up an upward vapor force, $F_v$ (N). If $F_v$ is greater than the droplet's weight, $G$ (N), it tends to lift it, whereas the liquid–solid adhesive force, $F_{Ad}$ (N), impedes this motion (**Fig. 3d**). Conversely, if $F_v < G$, the droplet tends to sink into the microtexture, which the adhesive force, $F_{Ad}$, also counters (**Fig. 3e**). In the concave configuration, the vapor has adequate space to escape; thus, $F_v$ drops (**Figs. 3f and 3h**). Subsequently, the droplet advances downward into the microtexture, assuming a near-flat convex shape (**Figs. 3g and 3i**) with $F_{Ad}$ acting upward, which indicates $F_v < G$.

To test the crucial role of $F_{Ad}$, we reduced the liquid–solid adhesion on silica with DRPs by functionalizing it with perfluorodecyltrichlorosilane (FDTS) (**Methods**), which yielded apparent CAs for water droplets in the range: $\theta_A = 164 \pm 2°$ and $\theta_R = 143 \pm 2°$. When we placed water droplets onto FDTS-coated DRPs, $T_L$ dropped to ~172°C (**Supplementary Materials, Fig. S5a, and Video S6**). The adhesion force between a DRP and water in our experimental configuration is given by $F_{Ad} = \gamma_{LV} \times \pi D \times (1 + \cos\theta_o)$ [41]. As the surface chemistry changed from silica to FDTS, the intrinsic CA ($\theta_o$) of water (on smooth and flat surfaces) increased from 40° for silica to 112° for FDTS, it caused a 64% reduction in $F_{Ad}$ and lowered $T_L$ significantly. This point demonstrates the sensitivity of the Leidenfrost phenomenon on adhesion rather than empirical cut-offs set for the apparent CAs of water droplets on them[21]. Particularly, not all superhydrophobic surfaces suppress the $T_L$ of water. But why did the onset of the Leidenfrost phenomenon on superhydrophobic silica with DRPs get delayed in comparison to that on superhydrophilic cylindrical pillars despite their identical surface chemistry and similar dimensions? The lower contact angle hysteresis in the former due to the robust entrapment of air should have led to a lower $T_L$ than that of the latter.

To address this question, first, we pinpoint some crucial differences in the ways DRPs interact with liquids in comparison to cylindrical pillars. The bioinspired DRP geometry is peculiar in the sense that despite the hydrophilicity of the substrate, it exhibits superhydrophobicity when liquids are placed on top because (i) the mushroom shape of the pillars prevents liquid imbibition so the droplet is in touch with the pillars' caps and air, and (ii) as the droplet recedes, it has to detach from individual pillars, which requires small energy barriers to surpass, affording low contact angle hysteresis[22, 35, 42, 43] (e.g., please, see the confocal images in **Fig. S5b-c** before and after FDTS coating on DRPs and their effect of L-S contact). In silica with DRPs, the pillar cap and base are comprised of silica (2 µm thickness) and silicon (50 µm), respectively. When the water droplet lands onto DRPs, it only touches the pillars for a depth of ~2 µm (**Fig. 4a**), after which the air–water interface assumes a convex shape and any further liquid movement does not increase the liquid–solid contact (**Fig. 4b**). Particularly, the liquid–solid contact on the DRPs is sensitive only to the dimensions of the pillars' caps and not to the penetration depth ($L_p$); this peculiarity would cease to exist at ultrahigh Weber numbers, where the drops would penetrate the microtexture, and those cases are beyond the scope of this study. In stark contrast to DRPs, when a water droplet rests on superhydrophilic cylindrical pillars (**Supplementary Materials, Fig. S6**, and **Video S7**), it tends to penetrate them (**Fig. 4c**), thus increasing the liquid–solid contact area and deepening the liquid penetration depth (**Fig. 4d**). The enhanced penetration depth remarkably impacts heat transfer, which we next investigate computationally by simulating the temperature distributions in DRPs and cylindrical pillars, followed by presenting an analytical model to derive expressions for $F_v$ to gain mechanistic insights into the onset of the Leidenfrost phenomenon.

**Computational and analytical modeling**

Temperature distributions in silica surfaces adorned with DRPs and cylindrical pillars as a function of water penetration depth were simulated using COMSOL® (COMSOL Multiphysics, version 5.4[44]). In this approach, we assumed that (i) a quasi-steady state, (ii) the water temperature was 99°C[3], (iii) the convective heat transfer coefficient for the vapor inside microtextures was 20 W/(m$^2$·K)[31] (**Fig. S7**), and (iv) the evaporation at the air–water interface was considered as an outflow of heat flux. The experimentally observed penetration depths of water droplets into the microtextures comprising DRPs and cylindrical pillars were 2 and 20 μm, respectively (**Figs. 4a** and **4c**). Consequently, the top region of the DRPs (**Fig. 4f**) were significantly hotter than that of the cylindrical pillars (**Fig. 4h**), which explains why the heat transfer in the case of the DRPs was significantly lower than that in the case of cylindrical pillars (**Fig. 4e** and **4g**). The higher heat transfer in the case of cylindrical pillars generated higher vapor fluxes and hence the upward vapor force, thereby lifting the droplet at lower temperatures than that in the case of DRPs.

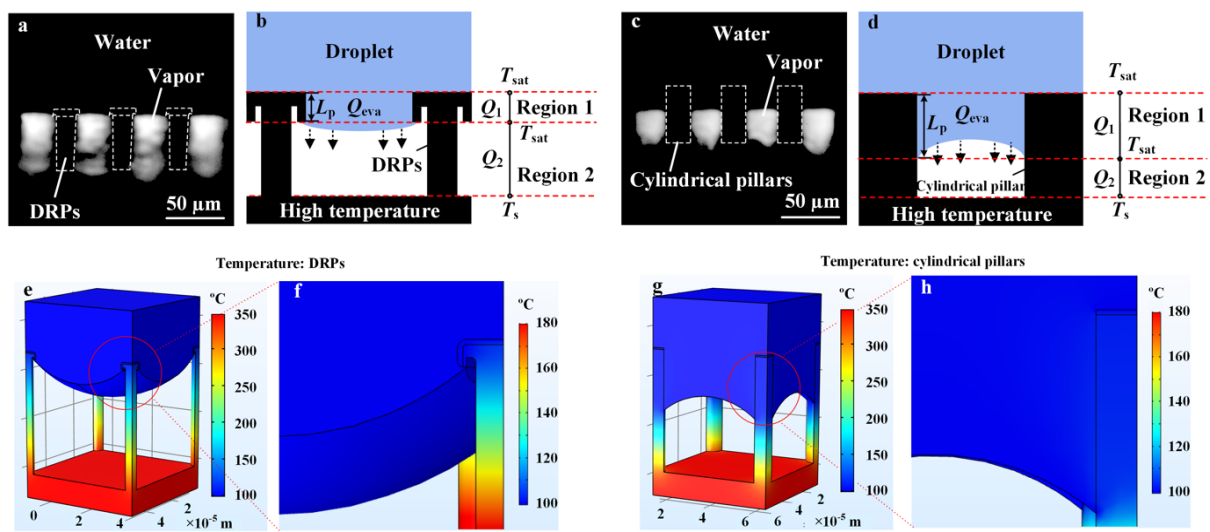

**Fig. 4| Heat transfer analysis at the liquid–solid interface as a function of penetration depth.** Representative experimental micrograph of a water droplet penetrating silica surfaces adorned with (a) DRPs and (c) cylindrical pillars of identical diameter, height, and pitch. Schematics of water droplet penetration into DRPs (b) and cylindrical pillars (d), pinpointing heat transfer rates. Here, $Q_{eva}$, is the overall evaporation heat; $Q_1$ is the heat transfer rate in Region 1; and $Q_2$ is the heat transfer rate in Region 2 (i.e., from the hot substrate). Temperature distributions due to heat transfer at the liquid–solid interface of (e, f) DRPs and (g, h) cylindrical pillars.

Next, we developed an analytical model to exploit these temperature profiles to estimate the vapor force, $F_v$. The assumptions underlying this quasi-steady state analytical model include (i) the temperature of the pillars in Region 1, wherein water penetrates, is the same as that of the water (based on simulation results in **Fig. 4e–h**), (ii) heat transfer is dominated by conduction and the contribution of convection is negligible[12], (iii) the flow of vapor across the microtexture is laminar; i.e., the characteristic Reynolds number is low, which is typical for the Leidenfrost phenomenon (at 100°C, the density of the water vapor is $\rho = 0.4$ kg/m³, the viscosity of the water vapor is $\mu = 2 \times 10^{-5}$ Pa·s, and the speed of the vapor in the microtexture is $U \approx 1$ m/s; thus, $Re = \rho_v U H/\mu$ is ~1)[3, 45]. The energy for water evaporation ($Q_{eva}$ (W)) is received through the heat influx at the water interface, $Q_1$ (W) (Region 1 in **Figs. 4b and 4d**). This heat transfer rate, $Q_1$, is also equal to the heat transfer rate absorbed from the substrate underneath $Q_2$ (W) (Region 2 in **Figs. 4b and 4d**). Therefore,

$$Q_{eva} = -\frac{dm_e}{dt} h_{fg} = Q_2 \qquad (1)$$

where $Q_2 = q_H A_o$, where $q_H$ (W/m²) is the heat flux from the hot surface. $A_0 = L^2$ (m²) is the projected area of the periodic unit (**Fig. 5a**), $m_e$ (kg) is the mass of evaporated water in a periodic cell, $t$ (s) is time, and $h_{fg}$ (J/kg) is the latent heat of the evaporation of water. Accounting for heat conduction across the vertical direction, Eq. (1) yields

$$\rho_v u_z A_v h_{fg} = k_{eff} \frac{dT}{dz} A_o \qquad (2)$$

where $\rho_v$ (kg/m³) is the density of the water vapor, $u_z$ (m/s) is the vapor velocity along the z-axis, $A_v = L^2 - \pi(D/2)^2$ (m²) is the vapor escape area inside the periodic cell, and $T$ (°C) is the temperature. The thermal gradient, $dT/dz$ (°C/m), depends on the penetration depth of water into the microtexture. Considering the droplet keeps moving up and down during heating (**Fig. 5a**), the average thermal gradient, $dT/dz$ was taken as $(T_s - T_{sat})/(H - L_p/2)$ in the calculation. Here, $T_s$ (°C) and $T_{sat}$ (°C) are the surface temperature of substrates and the boiling point of water, respectively; $H$ (m) is the height of DRPs, and $L_p$ is the penetration depth when the droplet moves down, ranging within, $0 < L_p < H$.

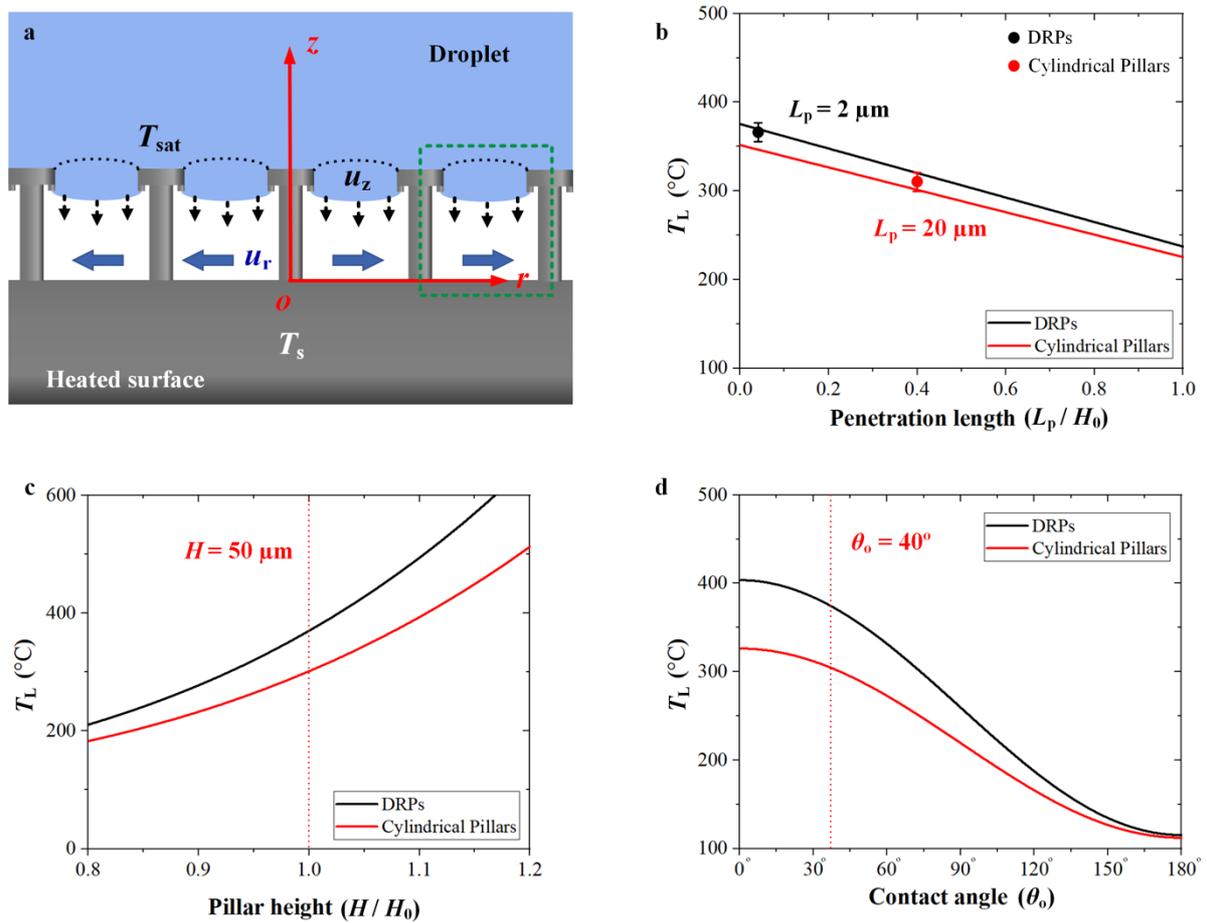

**Fig. 5| Water $T_L$ as a function of penetration depth, pillar height, and contact angle.** a) Schematic representing the flow of water vapor escaping through the microtexture. b) Water $T_L$ as a function of water penetration depth (rendered dimensionless by dividing it with the pillars' height, $H_0 = 50$ μm). c) Water $T_L$ as a function of pillar height (rendered dimensionless by dividing it with $H_0$). d) Water $T_L$ as a function of the actual (or intrinsic) contact angle.

Note that $k_{eff}$ (W/(m·K)) is the effective heat transfer coefficient determined by the area average of air and the solid fractions[17]:

$$k_{eff} = \varepsilon k_v + (1-\varepsilon)k_s \tag{3}$$

where $k_v$ (W/(m·K)) and $k_s$ (W/(m·K)) are the thermal conductivities of the vapor and the solid, respectively, and $\varepsilon$ is the ratio of the vapor area to the projected area, $\varepsilon = (L^2 - \pi(D/2)^2)/L^2$. Since the radius of the droplet's base ($R_b$ (m), ~0.7 mm from experimental observation) is dramatically larger than the height of the pillars, i.e., $R_b \gg H$, the lubrication analysis can be applied[46]. As the vapor is produced from the bottom of the droplet and released from the spaces between pillars, the continuity equation can be written as

$$\int_0^H u_r 2\pi r dz = \pi r^2 u_z, \tag{4}$$

where $u_r$ (m/s) is the radial vapor velocity. Considering an approximately linear temperature distribution along the pillars[12], we can obtain the vapor velocity along the z-axis:

$$u_z \approx \frac{k_{eff}(T_s - T_{sat})}{\varepsilon \rho_v h_{fg}(H - L_p/2)} \tag{5}$$

Thus, the average vapor velocity at the r-direction ($\overline{u_r}$) can be obtained:

$$\overline{u_r} = \frac{1}{H}\int_0^H u_r dz = \frac{ru_z}{2H} = r\frac{k_{eff}(T_s - T_{sat})}{2\varepsilon \rho_v h_{fg} H(H - L_p/2)} \tag{6}$$

After applying lubrication analysis, the momentum equation for flow in the porous media can be reduced to classical Brinkman Equation[47]:

$$\mu_v \frac{\partial^2 u_r}{\partial z^2} = \varepsilon \frac{dP_r}{dr} + \frac{\mu_v \varepsilon}{K} u_r \tag{7}$$

where $P_r$ (Pa) is the pressure distribution at the r-direction, $\mu_v$ (Pa·s) is the dynamic viscosity of vapor, and $K$ (m²) is the permeability of the porous media[48]. We apply the non-slip boundary conditions and estimate the average radial velocity as

$$\overline{u_r} = \frac{1}{H}\int_0^H u_r dz = \frac{K}{\mu_v}\left(\frac{dP_r}{dr}\right)\left(\frac{\tanh(H\beta/2)}{H\beta/2} - 1\right) \tag{8}$$

where $\beta = \sqrt{\varepsilon/K}$. Next, we substitute Eq. (6) into Eq. (8), and applying the boundary conditions, we obtain the radial pressure profile as

$$P_r = (R_b^2 - r^2)\frac{\mu_v k_{eff}(T_s - T_{sat})}{4\varepsilon\rho_v h_{fg} H(H - L_p/2)K}\left(1 - \frac{\tanh(H\beta/2)}{H\beta/2}\right)^{-1}. \tag{9}$$

Thus, the vapor force, $F_v$ (N), exerted by this vapor film onto the droplet can be estimated by

$$F_v = \int_0^{R_b} P_r 2\pi r dr = \pi R_b^4 \frac{\mu_v k_{eff}(T_s - T_{sat})}{8\varepsilon\rho_v h_{fg} H(H - L_p/2)K}\left(1 - \frac{\tanh(H\beta/2)}{H\beta/2}\right)^{-1}. \tag{10}$$

Furthermore, the meniscus adhesive force on each DRP, $F_{adi}$ (N), is estimated as

$$F_{ad_i} = \pi(D + 2S_h)\gamma_{LV}(1 + \cos\theta_o), \tag{11}$$

where $D$ (m) is the diameter of the top surface of DRPs, $S_h$ (m) is the horizontal length of edge DRPs, $\gamma_{LV}$ (N/m) is the surface tension of water, and $\theta_o$ is the actual (or intrinsic) contact angle of water on the silica surface. The Leidenfrost phenomenon occurs when the vapor force ($F_v$) overcomes the adhesive force ($F_{ad}$) and gravitation force ($G$), which allows the droplet to be levitated by a continuous vapor layer:

$$F_v = G + \sum_{i=1}^n F_{ad_i}, \tag{12}$$

where $G = mg$ is the gravitation force, $m$ (kg) is the mass of the droplet, $g$ (m/s²) is the gravitation acceleration, and $n$ is the number of DRPs in contact with the water droplet and can be estimated as $n = \frac{\pi R_b^2}{L^2}$. At the onset of the Leidenfrost phenomenon, $T_s$ is equal to $T_L$. Substituting Eqs. (10) and (11) into Eq. (12), the temperature corresponding to $T_L$ is

$$T_L - T_{sat} = \frac{8\varepsilon\rho_v h_{fg} H(H - L_p/2)K}{\pi R_b^4 \mu_v k_{eff}} \times \left(\left(\frac{\pi R_b^2}{L^2}\right)\pi(D + 2S_h)\gamma_{LV}\cos\theta_r + G\right) \times \left(1 - \frac{\tanh(H\beta/2)}{H\beta/2}\right) \tag{13}$$

Equation (13) captures the dependence of the $T_L$ on the thermal, structural, chemical, and inertial characteristics of the system, especially the heat conduction at the water–substrate interface that scales with the penetration length ($L_p$), microtexture dimensions, liquid–solid adhesion, and the vapor flow through the microtexture. As the penetration depth, $L_p$, increases, $T_L$ drops for both microtextures (**Fig. 5b**). To disentangle the effects of geometry and surface chemistry, we utilized the model to predict the variation in the $T_L$ of water with pillar height and surface chemistry. We found that $T_L$ in both cases increased with increased pillar height (**Fig. 5c**) because the taller the pillars are for a given pitch, the larger is the space to escape for the vapor, lowering $F_v$. The model also pinpointed that as the surface hydrophobicity increases, $T_L$ drops (**Fig. 5d**). Therefore, the onset of the Leidenfrost phenomenon depends on a delicate balance among $F_v$, $G$, and $F_{Ad}$, which cannot be satisfied by all superhydrophobic surfaces. Thus, these forces could be balanced to engineer superhydrophobic surfaces that can prevent the onset of the Leidenfrost phenomenon even better than hydrophilic and superhydrophilic surfaces.

**Discussion**

These experimental and theoretical findings might aid the rational design of superhydrophobic surfaces in thermal power engineering, for instance, for reducing frictional drag without compromising interfacial heat transfer. We illustrate this by comparing the time it takes to evaporate a drop of water placed on silica with DRPs with that on Glaco-coated silica, both maintained at 200°C. Although both surfaces were superhydrophobic, the water evaporated much faster on silica with DRPs than that on Glaco-coated silica (**Fig. 6**). Silica with DRPs facilitated a 300% higher heat transfer than Glaco-coated silica because the former suppressed the Leidenfrost phenomenon to well above 200°C, whereas the latter experienced film boiling at $T_L \approx 130°C$.

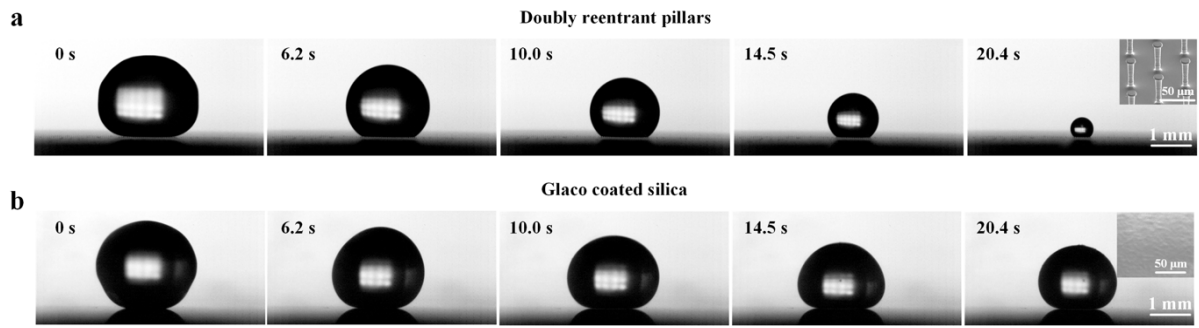

**Fig. 6| Evaporation of water droplets on superhydrophobic surfaces at a surface temperature of 200°C.** a) Silica with DRPs. b) Glaco-coated silica.

Penetration depth was recognized as a crucial factor influencing the water–substrate heat transfer, and for a given microtexture, it depends on the Weber number[49]. At low Weber numbers, such as in the experiments reported here, DRPs pinned the liquid meniscus at the doubly reentrant edge such that any further penetration did not increase the liquid–solid contact area. The DRP architecture helped us disentangle the effects of the penetration depth and the droplet–substrate contact area on the onset of the Leidenfrost phenomenon. However, at higher Weber numbers, the liquid meniscus might de-pin from doubly reentrant edges and land onto the DRPs' stems[50]. Subsequently, akin to the case of cylindrical pillars, the curvature of the liquid–vapor interface will reverse, the liquid-solid contact area will increase, and the conductive heat transfer will yield the Leidenfrost phenomenon. However, as the droplet's momentum dissipates and once it touches the solid bottom surface, DRPs' original behavior, suppressing the Leidenfrost phenomenon, might get restored. These predictions necessitate a systematic investigation. Notably, the coating-free aspect of the DRP architecture would offer resilience against harsh operational and cleaning protocols and organic fouling[51] as perfluorinated coatings/materials are vulnerable to these conditions[52]. In fact, DRP-based coating-free design principles are being exploited for mitigating cavitation[53] and water desalination via scalable manufacturing platforms[54].

In conclusion, we have demonstrated that it is possible to realize superhydrophobic surfaces that do not suppress the $T_L$ of water via a judicious choice of surface microtexture and chemistry. This design can yield superhydrophobic surfaces or coatings for simultaneously reducing frictional drag and preempting boiling crisis in thermal machinery, such as heat exchangers, boiler pipes, and turbine blades. The coating-free aspect of the bioinspired DRP architecture is also enticing for its robustness against harsh conditions and organic fouling.

## Methods

*Sample preparation*

Silicon wafers (p-doped, 4-inch diameter, and 500 μm thickness with a 2.4-μm-thick thermally grown silica layer) were used as substrates (**Fig. 1a**). To fabricate DRPs (**Fig. 1d**), we applied photolithography, dry etching, and thermal oxide growth, followed by the protocols reported in the previous studies.[23, 37] Besides DRPs, we microfabricated silica surface adorned with arrays of cylindrical pillars, which were rendered superhydrophilic (**Fig. 1c**), manufactured by the same materials above, and the pitch, height, and top diameter of cylindrical pillars were the same as those of the DRPs. After surface fabrications and before droplet experiments, all samples were cleaned with Piranha solution ($H_2SO_4$: $H_2O_2$ = 3:1) for 15 min at 110°C and stored in the vacuum oven for 24 h at 50°C and ~0.03 bar. Consequently, the actual (or intrinsic) contact angle ($\theta_o$) of water on the silica surface was stabilized at ~40°.

For comparison, Glaco coating was used to coat flat silica (same materials in the DRPs) as the common superhydrophobic surface (**Fig. 1b**). The effect of surface chemistry was studied by coating perfluorodecyltrichlorosilane (FDTS) on DRPs using a molecular vapor deposition system (ASMT 100E), as described previously[53]. All the surface structures were examined by a scanning electron microscope (Helios Nanolab SEM).

*Contact angle measurements*

We characterized the wettability of these surfaces with water by measuring apparent CAs of water droplets (6 μL) and advancing and receding angles at a rate of 0.2 μL/s. The water CAs on the surfaces of different samples were measured using a drop shape analyzer (Kruss: DSA100) interfaced with the Advance software. The sliding angles of water droplets (10 μL) on these surfaces were measured by an automatic tilting stage, where a camera was used to record the movement of droplets. All experiments were repeated at least three times.

*Droplet generation for Leidenfrost experiments*

To generate water droplets, we used a Harvard Apparatus syringe pump (PHD ULTRA: 703007INT) and a 60-mL plastic syringe connected by a plastic syringe extension tube. A stainless-steel capillary (diameter: ~0.51 mm) was used to dispense water perpendicular to heated test surfaces. The flow speed was set as 100 μL/s, and the syringe system continuously generated 15-μL droplets (radius: ~1.5 mm, which is smaller than the capillary length of water ~2.7 mm at NTP).

*Confocal experiments*

A Zeiss upright laser confocal microscope (Model number: LSM710) was used to detect the meniscus at the interface between a droplet and pillars. A ~100-μL water droplet with Rhodamine B (Acros) was gently placed on samples using a manual syringe, and a 20X micro-objective lens was immersed into the droplet before acquisition. The 520-nm laser at the Z-stack mode was used to scan the droplet–surface interface. Subsequently, using Imaris v.8.1 by Bitplane, we obtained cross-sectional images to visualize the meniscus shape (**Supplementary Materials, Fig. S5b-c**).

*High-speed imaging*

A high-speed camera (model number: Phantom V1212) with a zoomed lens was applied to record images at a high frame rate. To ensure that all the images have the same configuration, the imaging parameters of the camera such as frame per second (FPS), image pixels, and exposure time in all experiments were set to 1000 FPS, 512 × 512 pixels, and 200 μs, respectively. A triaxial stage was used to adjust the position of the samples with respect to the high-speed camera. The hot plate was leveled using a bubble leveler, and the images were calibrated by the stainless-steel capillary. For image processing, Phantom Camera Control 2.8 software was used.

*Leidenfrost experiments*
A hot plate (temperature: 20°C–540°C) was placed on a triaxial stage and balanced by a bubble level. Considering the nonequilibrium nature of the hot plate, the actual temperatures of the hot plate surface were measured by a K-type thermocouple and recorded by a temperature data logger (Omega: RDXL6SD).

We moved the capillary tip with a droplet as close to the surfaces as possible to gently release droplets from the capillary onto the surfaces to minimize the droplet impact. Thus, all the droplets keep the same low Weber number (***We*** ≈ 1.7) to minimize its effect on the Leidenfrost point. Water droplets (15 μL) were generated by a syringe pump and released from the needle tip to the heated sample surface. A high-speed camera was used to record the impact, bounce, and vibration of droplets and observe liquid–vapor–solid interactions at the interface between droplets and surfaces. All the experiments were repeated at least three times.

To measure Leidenfrost points on different surfaces, the classical lifetime method[10] was used to measure $T_L$ on these surfaces. The first peak in the curve between the lifetime and surface temperatures yields $T_L$[10, 25]. To measure the lifetimes of droplets, the temperature of the hot plate increased by 10°C at each step, and the droplet evaporation times were recorded by a timer. All experiments were repeated at least five times. Since there is usually no clear lifetime peak when the Leidenfrost transition occurs on superhydrophobic surfaces, we used the high-speed camera to detect $T_L$ on Glaco-coated flat silica and silica with DRPs by observing the moments of vapor film formation and droplet bouncing while gradually increasing the temperature.

*Calculations*
Simulation results were obtained by utilizing the heat transfer module in COMSOL (5.4) (the detailed boundary conditions are shown in **Supplementary Materials**). Equations 1–13 were solved using the commercial software Wolfram Mathematica (11.0). The thermal properties of water were taken at 99°C, and other materials' properties were taken at 225°C, which is an average temperature between the water and the hot substrate.

**Acknowledgments**
This work was supported by King Abdullah University of Science and Technology (KAUST). M.S. thanks Prof. Sigurdur Thoroddsen from KAUST and Prof. Shangsheng Feng from Xi'an Jiaotong University for fruitful discussions.

**Supplemental Information**

**Unexpected Suppression of Leidenfrost Phenomenon on Superhydrophobic Surface**

Meng Shi, Ratul Das, Sankara Arunachalam, Himanshu Mishra*

Interfacial Lab (iLab), Water Desalination and Reuse Center (WDRC),

Biological and Environmental Science and Engineering (BESE) Division,

King Abdullah University of Science and Technology (KAUST),

Thuwal, 23955-6900, Saudi Arabia

*Corresponding author: himanshu.mishra@kaust.edu.sa

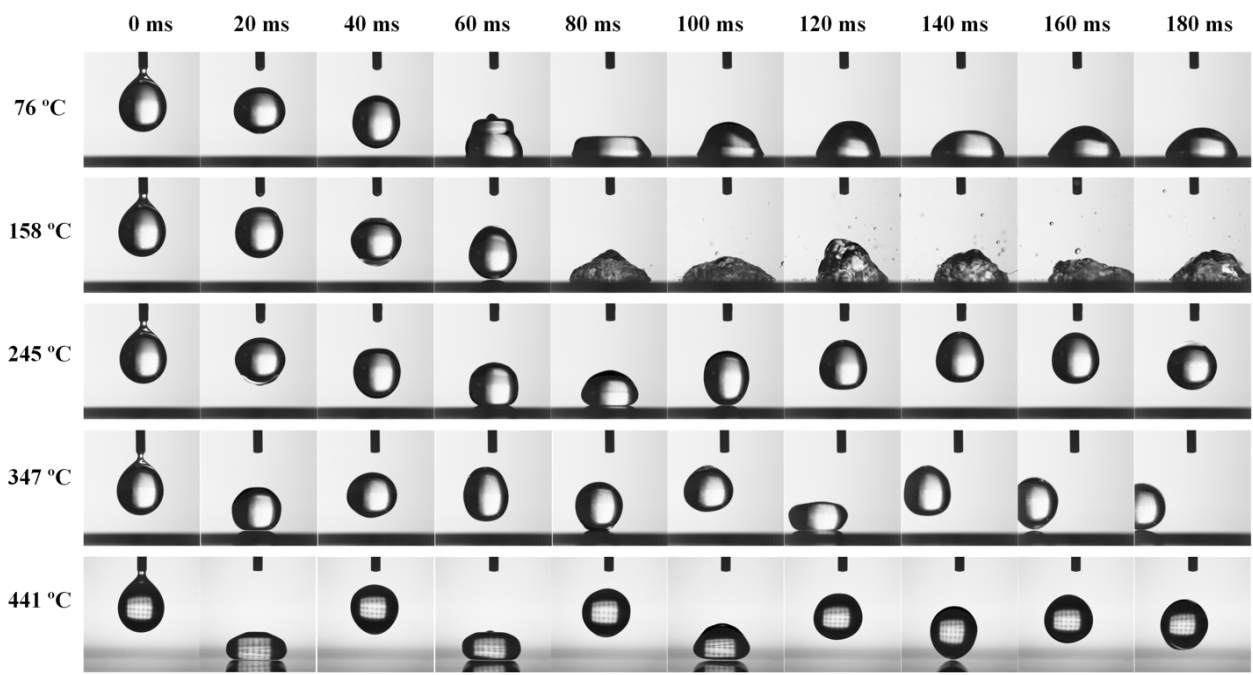

**Figure S1 Droplets' behaviors on flat silica surfaces maintained at different temperatures:** the Leidenfrost phenomenon appeared at $T_L \approx 286°C$. Representative snapshots of droplet behavior when the surface temperature was below $T_L$ (76°C, 158°C, and 245°C) and above $T_L$ (347°C and 441°C). Although the droplets bounced after initially contacting the hot surface at 245°C, we observed transition boiling at the bottom of the droplet after several bounces. Therefore, the droplet is in transition boiling at 245°C, which can also be seen from the lifetime curve (**Figure 2e**).

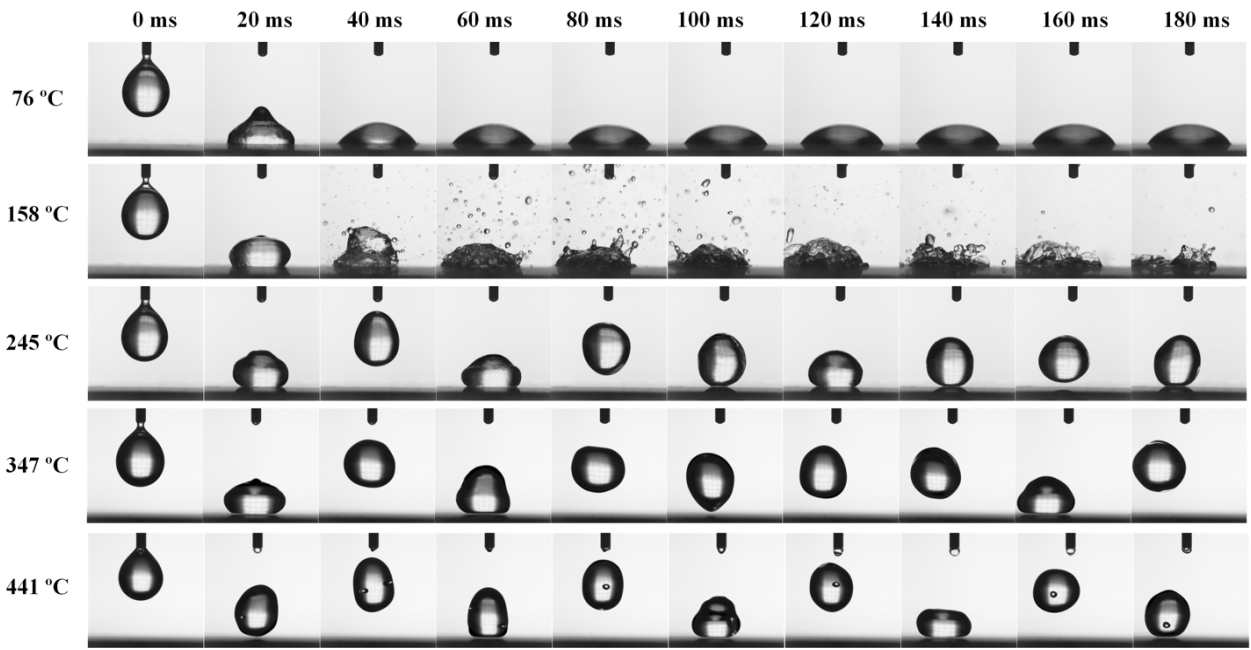

**Figure S2 Droplets' behaviors on silica with cylindrical pillars maintained at different temperatures:** the Leidenfrost phenomenon appeared at $T_L \approx 314°C$. Representative snapshots of droplet behavior when the surface temperature was below $T_L$ (76°C, 158°C, and 245°C) and above $T_L$ (347°C and 441°C). Although droplets bounced when initially contacting the hot surface at 245°C, we observed transition boiling at the bottom of the droplet after several bounces. Hence, the droplet is in transition boiling at 245°C, which can also be deduced from lifetime curve (**Figure 2e**).

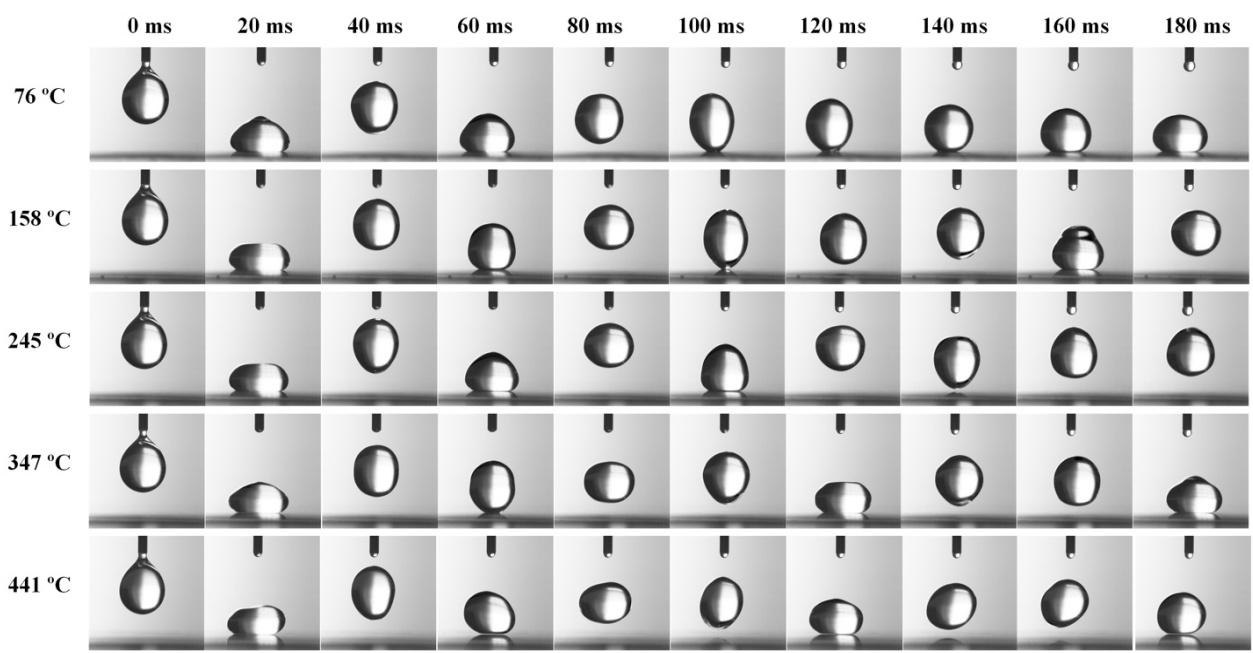

**Figure S3 Droplets' behaviors on Glaco-coated flat silica maintained at different temperatures:** the Leidenfrost phenomenon appeared at $T_L \approx 130°C$. Representative snapshots of droplet behavior when the surface temperature was below $T_L$ (76°C) and above $T_L$ (158°C, 245°C, 347°C, and 441°C).

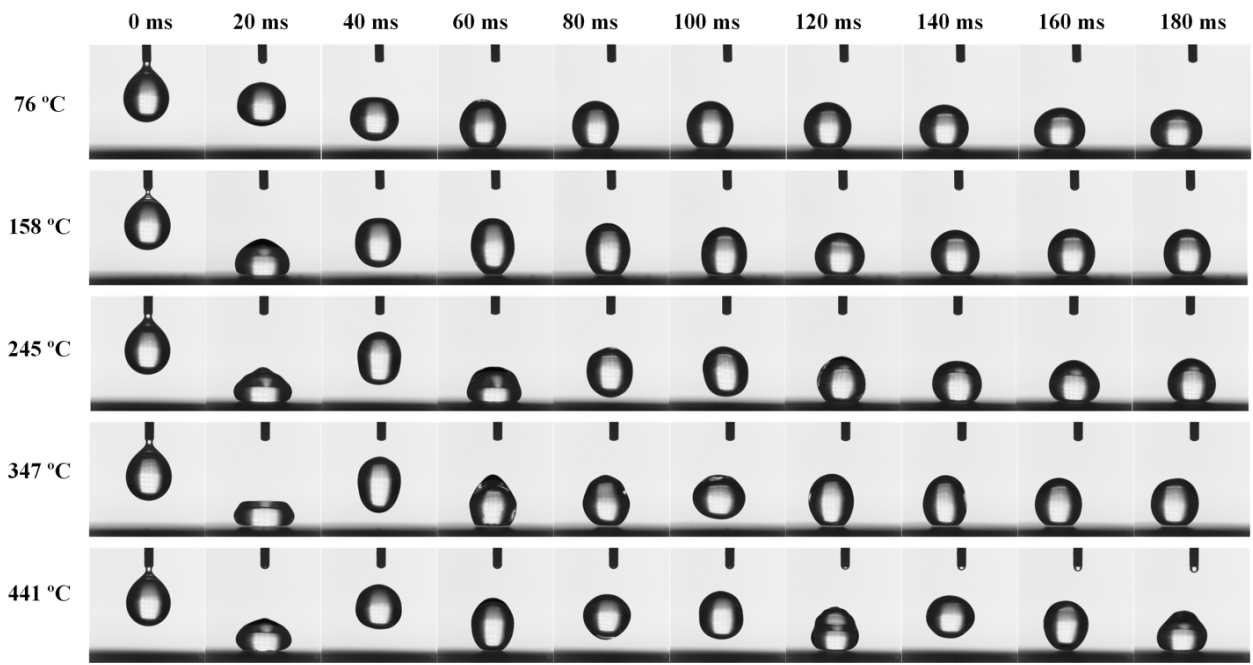

**Figure S4 Droplets' behaviors on silica with DRPs maintained at different temperatures:** the Leidenfrost phenomenon appeared at $T_L \approx 363°C$. Representative snapshots of droplet behavior when the surface temperature was below $T_L$ (76°C, 158°C, 245°C, and 347°C) and above $T_L$ (441°C).

We performed droplet experiments on the FDTS-coated DRPs at the same temperature as those performed on other surfaces (i.e., ~347°C) and found that the water droplet was levitated by a vapor film and started bouncing on the FDTS-coated DRPs (**Figure S5a**). Through confocal experiments, we obtained water droplet meniscus curvatures at the interfaces of FDTS-coated and untreated DRPs. On the untreated DRPs, a water droplet meniscus goes down to the doubly reentrant edges (**Figure S5b**). After FDTS coating, a water droplet can stay only on the top of the DRPs and cannot wet the edges (**Figure S5c**). Thus, the adhesion is significantly reduced by not only increasing the actual (intrinsic) contact angle $\theta_o$ to 112° but also changing the contact length slightly. Thus, the local pinning effect is eliminated from the surface of the FDTS-coated DRPs.

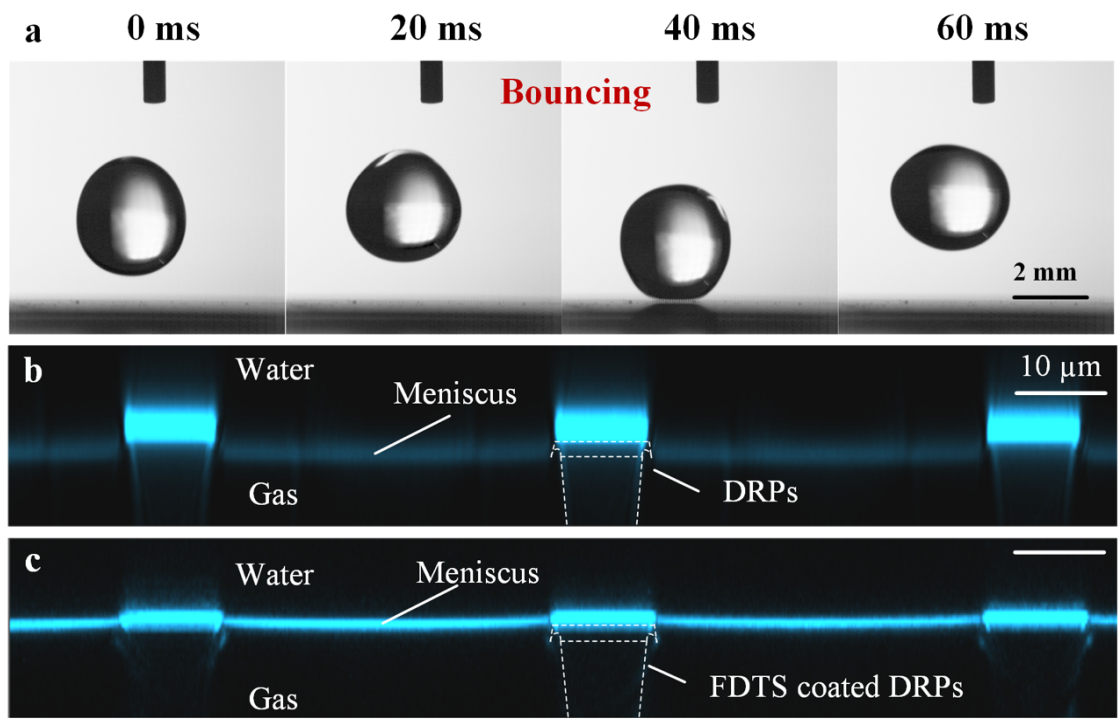

**Figure S5 Effect of hydrophobic coating on the Leidenfrost phenomenon on silica with doubly reentrant pillars.** a) Droplet bouncing on FDTS-coated DRPs at 347°C; b) Droplet meniscus at the interface between the droplet bottom and DRPs; c) Droplet meniscus at the interface between the droplet bottom and FDTS-coated DRPs.

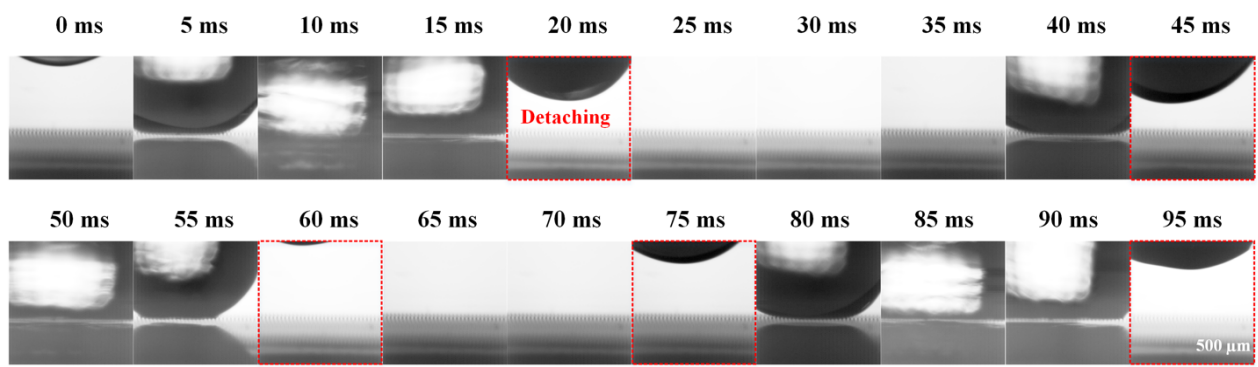

**Figure S6 Droplet detachment from cylindrical pillars at 347°C.** Once a water droplet touches the surface of the heated pillars, it deforms and penetrates the spaces between the pillars. Then, the water will absorb heat from the pillars and evaporate rapidly. The generated vapor will push the water upward. After the water detaches from the tops of all the simple pillars, the droplet will be levitated by the generated vapor.

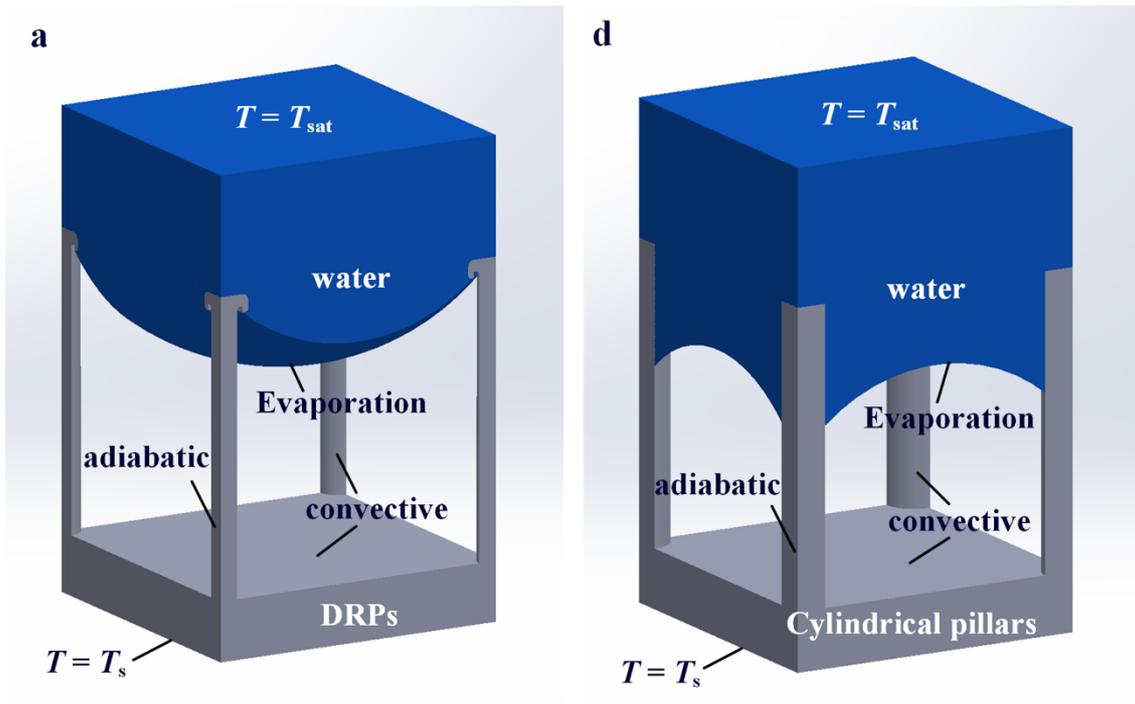

**Figure S7 Geometrical models and boundary conditions of heat transfer simulations**

**Supplemental Videos**

**Video S1 Droplets bouncing on the smooth silica surface at 347°C**

**Video S2 Droplets bouncing on the Glaco-coated smooth silica surface at 347°C**

**Video S3 Droplets bouncing on simple pillars at 347°C**

**Video S4 Droplets sitting on doubly reentrant pillars at 347°C**

**Video S5 Interface of a water droplet on doubly reentrant pillars at higher magnification at 347°C**

**Video S6 Droplets bouncing on FDTS-coated doubly reentrant pillars at 347°C**

**Video S7 Droplet detachment on simple pillars at 347°C**